\documentstyle[aps,epsf,pre,twocolumn]{revtex}

\draft 

\begin{document}

\title{Temporal Modulation of the Control Parameter
in Electroconvection in the Nematic Liquid Crystal I52}

\author{Michael Dennin}
\address{Department of Physics and Astronomy}
\address{University of California at Irvine}
\address{Irvine, CA 92697-4575.}

\date{\today}

\maketitle

\begin{abstract}

I report on the effects of a periodic modulation 
of the control parameter on electroconvection
in the nematic liquid crystal I52. Without modulation,
the primary bifurcation from the uniform state is a direct
transition to a state of spatiotemporal chaos.
This state is the result of the interaction of four, degenerate
traveling modes: right and left zig and zag rolls.
Periodic modulations of the driving voltage
at approximately twice the traveling frequency are used. 
For a large enough modulation amplitude,
standing waves that consist of only zig or
zag rolls are stabilized. The standing waves exhibit regular
behavior in space and time. Therefore, modulation of the control
parameter represents a method of eliminating spatiotemporal
chaos. As the modulation frequency is varied
away from twice the traveling frequency, standing waves that are
a superposition of zig and zag rolls, i.e. standing rectangles,
are observed. These results
are compared with existing predictions based on coupled
complex Ginzburg-Landau equations.
\end{abstract}


\section{Introduction}

When a spatially extended system is driven far from
equilibrium, a series of transitions occurs
as a function of the external driving force,
or control parameter \cite{REV}. The initial transition is typically
from a spatially uniform state to a state with periodic
spatial variations, called a pattern. As the control parameter is
increased, a sequence of instabilities occurs that produces
increasingly complex spatiotemporal patterns. Ultimately, the system
becomes fully turbulent.
States of spatiotemporal
chaos form an interesting
class of patterns \cite{REV,EXP}.
Loosely speaking, spatiotemporal chaos refers to
deterministic patterns that possess
a random variation in space and time. They
generally exhibit an underlying
periodicity but are characterized by a
finite correlation length and correlation time. Since
their discovery in fluid dynamical systems \cite{DISC},
states of spatiotemporal chaos have been observed
in a wide range of systems, including
Rayleigh-B\'{e}nard convection, Faraday instabilities,
Taylor-Couette flow, electroconvection in nematic
liquid crystals, lasers, and chemical reactions  \cite{REV,EXP}.
Despite the ubiquitous nature of the phenomenon, a full
understanding of spatiotemporal chaos remains
one of the outstanding problems in nonlinear dynamics.

One of the challenges facing the study of spatiotemporal
chaos is the difficulty associated with characterizing
the dynamics \cite{REV}. Because the systems are spatially extended,
they are inherently high dimensional. This precludes the
use of many of the successful techniques developed to
study low dimensional chaos in dynamical systems \cite{CHAOS}.
Also, most examples of spatiotemporal chaos occur at sufficiently
large values of the control parameter that weakly nonlinear
techniques, such as amplitude equations, are only
applicable as phenomological models, if at all. The work
reported here uses a state of spatiotemporal chaos that
occurs in electroconvection in I52 \cite{DAC96}. In contrast to other
systems, this state does occur in the weakly nonlinear regime
where quantitative comparison between theory
and experiment is possible. A fundamental theory
of electroconvection, the WEM model \cite{TK95,DTKAC96}, exists
and provides the necessary starting point for a
{\it quantitative} derivation of the relevant
amplitude equations \cite{TK98}. Therefore, this system is an ideal
candidate
for studying spatiotemporal chaos. However, currently only
two of the four
necessary amplitude equations have been derived and only
qualitative predictions exist \cite{TK98}.
In this work, I use temporal modulation as a probe of the system's
dynamics. This provides both a test
of the existing amplitude equation description and a means
of guiding future experiments and calculations.

For electroconvection \cite{SMREFS,ECART},
a nematic liquid crystal
is placed between two properly treated glass plates
so that the director is everywhere parallel to the
plates and along a chosen axis. A nematic
liquid crystal is a fluid in which the molecules have
orientational order, and the director refers to the axis
parallel to the average alignment of the molecules \cite{LC}.
The liquid crystal is doped with ionic impurities,
and an ac voltage is applied across the sample using
transparent electrodes on the glass plates.
Above a critical value of the applied voltage,
there exists a transition from uniform conduction to
convection rolls, with a corresponding periodic variation of
the director and concentration of ionic impurities.
Because the system is anisotropic, the patterns
can be characterized by the angle $\theta$
between the director and the wavevector of
the pattern.
Electroconvection in the nematic
liquid crystal I52 exhibits
a forward, Hopf bifurcation to oblique rolls \cite{DAC96}.
A Hopf bifurcation is a transition to a traveling
wave pattern, and oblique rolls correspond to patterns
where $0^\circ < \theta < 90^\circ$ \cite{SMREFS}.
At onset, because the director only defines an axis
and not a positive direction, states with
wavevectors of equal magnitude but with angle's
$\theta$ and $\pi-\theta$ are degenerate
and referred to as zig and zag rolls, respectively.
The state close to onset that is studied here consists
of four modes: right- and left-traveling zig 
and zag rolls \cite{DAC96}. It is the interaction of these
four modes that results in spatiotemporal chaos.

Resonant modulation of the control parameter in a system
with a Hopf bifurcation is known to stabilize standing waves for large
enough modulation strength \cite{RCK88,W88,RRFJS88,JZR89,JR90,RSK94}.
However, it has not been applied previously to a state
of spatiotemporal chaos. There are three main questions
addressed in this work. First, I have made
a survey of the range of existence
of standing wave patterns and their qualitative features
for a reasonable set of values of the
applied voltage, applied frequency, the modulation
strength, and the modulation frequency. Second,
there are three possible standing wave solutions:
standing rolls (only zig or zag rolls are present),
standing rectangles (superimposed zig and zag rolls
with equal amplitude), and standing
cross rolls (zig and zag rolls with different amplitudes) \cite{RSK94}.
I will show that the standing
wave states are generally standing rolls, but that standing rectangles
can be observed. Finally, I will demonstrate that the standing roll states
are temporally regular and that they eventually become spatially uniform.
Therefore, this is an example of the elimination of spatiotemporal chaos
and provides an interesting contrast to systems where temporal modulation
produces irregular behavior in an otherwise regular system \cite{JZR89,GB78}.

The rest of the paper is organized as follows. Section II provides
the details of the experimental techniques. Section III presents
the experimental results. In
this section, I will report separately on
the results of applying modulations
of the control parameter below and above the critical voltage
for the onset of convection in the absence of modulations.
Section IV will discuss the relationship between these
results and existing predictions of relevant amplitude equations.

\section{Experimental details}

The experiments were carried out using two electroconvection
cells containing the liquid crystal I52 \cite{FGWP89}.
The first cell was a custom made cell with a thickness of
25~${\rm \mu m}$. It was formed from two glass slides that were
coated with a layer of indium-tin oxide (ITO), a transparent conductor.
The conductive coating was etched
to form a 0.5~cm x 0.5~cm square electrode in the center of 
a 2.5~cm x 2.5~cm cell. 
The director was aligned using a rubbed polyimide.
The second cell
was a commercial cell obtained from EHC, LTD. \cite{EHCO}. This cell had
a thickness of 23~${\rm \mu m}$. The electrode was a
1.0~cm x 1.0~cm area of ITO in a 2.5~cm x 2.5~cm cell. The
alignment was also due to a rubbed polyimide coating. 
For both cells, there was some forcing of a pattern
due to fringing fields at the edge of the electrodes. However,
a region existed in the middle of each cell where convection
started spontaneously. All measurements were made
in this region of the cell to minimize boundary effects.

The main differences between the two samples were the values
of the critical voltage and the Hopf frequency due to differences in
iodine doping. The custom sample was filled with I52 that had been 
doped with $6.3\%$ by weight molecular ${\rm I_2}$ two months
prior to filling the cell. The commercial
cell was filled with I52 that had been doped with $6\%$ by weight
molecular ${\rm I_2}$ seventeen months prior to filling. The custom cell was
aged for five months after filling
before experiments were started, and the commercial
cell was aged for one month after filling.
For the custom cell, the critical voltage was approximately 21 V
and the Hopf frequency was approximately 0.125 Hz
at an applied frequency of 25 Hz.
For the commercial cell, the
critical voltage was approximately 15 V and the Hopf frequency was
approximately 0.34 Hz at an applied frequency of 25 Hz. 

It is known for electroconvection in I52 that the
level of iodine in the sample will drift in time. This results in drifts
in the critical voltage and the Hopf frequency. 
Because of this drift, the following protocols were used. The critical
voltage and Hopf frequency
were measured before and after every set of measurements.
The modulated voltage had the following form:
$V(t) = [V_o + V_m cos(\omega_m t)]cos(\Omega t)$.
The two main
dimensionless control
parameters are: $\epsilon = V_o^2/V_c^2 - 1$ and
$b = V_m/V_c$. Here $V_c$ is the critical voltage
for the onset of convection in
the absence of modulation. The drift in $V_c$ was
linear in time and corresponded to a drift in $\epsilon$ of
0.001/Hr. This drift is accounted for in all reported values of
$b$ and $\epsilon$. The modulation frequency will be discussed
in terms of the shift from resonance
$(f^* - f_m/2)$, where $f_m$ is the
modulation frequency and $f^*$ is the natural frequency
of the pattern. For $\epsilon < 0$, $f^*$ is the
Hopf frequency $f_h$, and for $\epsilon > 0$, it is the
frequency of the pattern in the absence of modulation.
Despite the differences between the two cells and the drift in time
of $V_c$, the behavior as a function of $b$, $\epsilon$,
and $(f^* - f_m/2)$ is completely
reproducible.
The sample temperature was held constant at either
$40 \pm 0.002\ ^\circ{\rm C}$ or $42 \pm 0.002\ ^\circ{\rm C}$. The
later temperature was used after the total drift in $V_c$ 
at $40 ^\circ{\rm C}$ had exceeded
approximately one volt.

Images were taken using a standard shadowgraph
method \cite{RHWR89} and are presented here with the
undistorted director
aligned in the horizontal direction. Because of the well-known
nonlinear effects of the shadowgraph method \cite{RHWR89},
the images have
been Fourier filtered so that only the fundamental modes
are present. This is essential for highlighting the standing
wave character of the modulated pattern.

The frequency of the pattern was determined by taking the Fourier
transform of a time series of 32 images. The images typically covered
a spatial area containing 18 rolls, though smaller regions of
only 7 rolls were also used. The time between images was chosen so
that the time series covered 4 to 6 periods of the fundamental
frequency.

For the measurements of the dynamics of the local amplitude of each mode,
a time series of 32 images was used. The time between images was chosen so
that the series consisted of four cycles of the pattern.
Each image covered a spatial area of approximately 7 wavelengths.
The modulus squared of the space-time Fourier transform
of the series,
$S({\bf k}, \omega)$, was used to compute the amplitude
of each mode.
The power in each mode, right- and left-traveling zig and
zag rolls, was determined by summing $S({\bf k}, \omega)$
over a 5 by 5 pixel grid in wavenumber space and 5 pixel window
in frequency space. The grid and window were
centered on the peak in $S({\bf k}, \omega)$ that corresponded
to the mode of interest.
The amplitude of each mode is the square root of the power.

For measurements of the onset of standing waves, $\epsilon$ was fixed and
the value of $b$ was either stepped up or down
in increments of 0.005. At each step,
the system was equilibrated for 10 minutes before a time series
of images was taken. The time series of images were used
to determine if the pattern was frequency locked to the modulation
and whether or not a standing wave had been established. 

\section{Experimental results}

\begin{figure}[htb]
\epsfxsize = 3.5in
\centerline{\epsffile{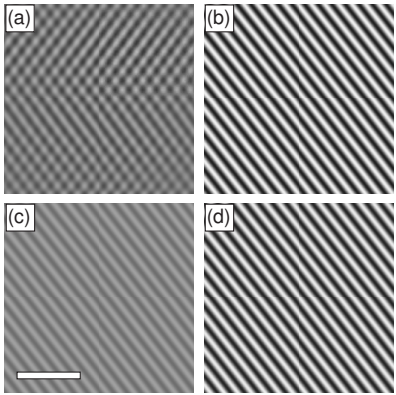}}
\caption{\label{space} Four images of the pattern in a
0.6~mm x 0.6~mm region of the cell. The
bar in (c) represent 0.2 mm. (a) An image of the cell
at $\epsilon = 0.03$ and no modulation. The image has been
Fourier filtered so that only the fundamental mode remains.
Image (b) through (d) are three images from a time series
taken at $\epsilon = 0.036$ and $b = 0.04$. The images
are 0.9~s apart and have also been Fourier filtered. The
modulation frequency was 0.694 Hz, which corresponds to
twice the Hopf frequency. These images illustrate the standing
wave nature of the pattern.}
\label{fig:space}
\end{figure}

Figure~1 is a comparison of the typical pattern in the region where
spatiotemporal chaos exists and the standing wave pattern that is
stabilized by the modulation. Figure~1a is a single snapshot of
the state of spatiotemporal chaos at a value of
$\epsilon = 0.03$ and no modulation. Figures~1b-d are three images
taken 0.9 s apart of the standing wave state at an $\epsilon = 0.036$,
$b = 0.04$, and $f_m = 2f_h$. The time between the images was chosen
to highlight the relative change in phase that is characteristic of
a standing
wave as one crosses the minimum in intensity. For example,
the intensities in
the lower right corner in Fig. 1b are opposite those in Fig. 1d.

\vskip 0.5in
\begin{figure}[htb]
\epsfxsize = 3.5in
\centerline{\epsffile{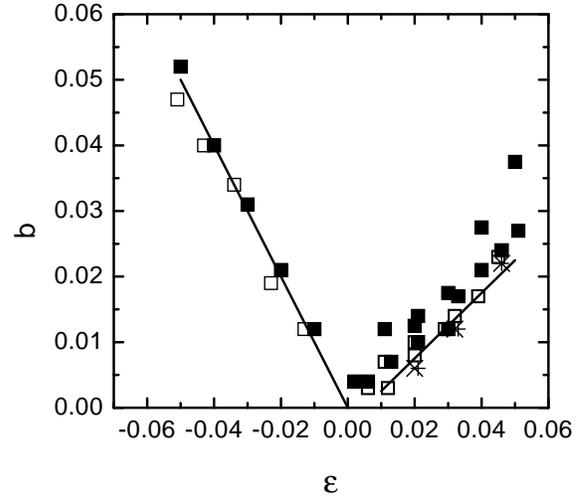}}
\caption{\label{onset} The symbols give the location of the
transition to standing
waves as a function of modulation strength $b$ and reduced
control parameter $\epsilon$. The solid symbols and
$\times$'s are
for increasing the value of $b$. The open symbols and
the $+$'s are
for decreasing the value of $b$. The difference between
the symbols is described in the text. For $\epsilon < 0$, there
is no pattern present below the symbols, and a spatially
uniform pattern exists above the symbols. For $\epsilon > 0$,
the state of spatiotemporal chaos exists below the symbols. Above
the symbols, the system is phase locked to the modulation, and
a state of standing rolls exists. For negative $\epsilon$, the
solid line is the expected onset based on coupled complex
Ginzburg-Landau equations, $b = \epsilon$.
For positive $\epsilon$, the solid
line is the curve $b = 0.57\epsilon - 0.006$ and provides a guide
to the eye for the transition values.}
\end{figure}

Figure 2 summarizes the range of existence of the standing wave patterns
when a modulation of twice the Hopf frequency is used.
The solid symbols and $\times$'s are the location of the
transition to standing waves
as measured by increasing $\epsilon$. The open symbols and
the $+$'s are the location of the transition to standing waves
as measured by decreasing $\epsilon$. In the region above and
between the two solid lines, the pattern is composed of uniform
standing rolls. As a check on the stability of the standing rolls,
two runs were made at fixed
$b$. For these runs, $\epsilon$ was increased in steps
of 0.005, with a waiting time of 5 minutes per step.
One run was at $b = 0.02$, and the other was
at $b = 0.05$. For the entire range of $\epsilon$ within the boundaries,
only uniform standing rolls were observed at these two values of
$b$.

For negative
values of $\epsilon$, the behavior of the system
is straightforward. The system
makes a transition directly from a uniform state to a state of 
frequency locked standing waves. The pattern consists of either
standing zig rolls or standing zag rolls, and never the superposition.
The pattern is also frequency locked to half the modulation frequency.
As shown in Fig. 2, within the resolution used here, there is
essentially no
hysteresis in the transition.

\vskip 0.5in
\begin{figure}[htb]
\epsfxsize = 3.5in
\centerline{\epsffile{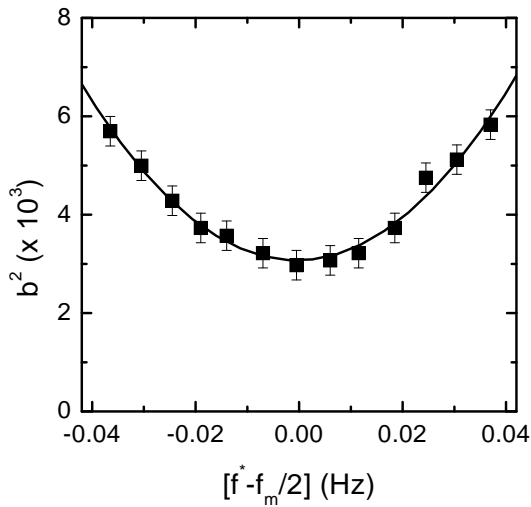}}
\caption{\label{negeps} Transition to standing waves for
$\epsilon = -0.054$. Below the curve, the system
is uniform. Above the curve, the state of the system is
standing rolls. The symbols are the experimental values. The
solid line is a fit to the expected curve:
$b^2 = \epsilon^2 + (2\pi\tau_d)^2(f^*-f_m/2)^2$. In this case,
$f^* = 0.162$, which is the Hopf frequency of the system.}
\end{figure}

Figure 3 shows the behavior of the system when the modulation frequency
is varied away from twice the Hopf frequency at negative $\epsilon$.
The solid squares represent the onset to standing waves, and the solid
line is a fit to a parabola. In this case, the parabola is centered on
the Hopf frequency.

For positive values of $\epsilon$, the situation is more
complicated because the ground state is the state of spatiotemporal chaos.
In this case, I have measured the transition in two different ways. First,
I have considered the onset to standing rolls. These are the solid and
open squares in Fig. 2. When stepping down in $\epsilon$, the transition
from standing rolls to the disordered state was easily identified. However,
because of the spatial disorder, when $\epsilon$ is increased,
the onset to spatially uniform
standing rolls is less well-defined. For the purposes of
Fig. 2, the onset was taken to be the point where the patches of
standing rolls had a size on the order of 18 wavelengths. 

\vskip 0.4in
\begin{figure}[htb]
\epsfxsize = 3.5in
\centerline{\epsffile{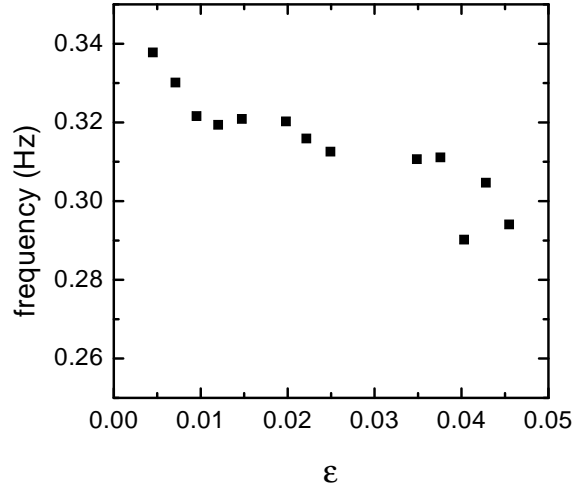}}
\caption{\label{freq} Measured frequency of the pattern
as a function of $\epsilon$. Illustrates the decrease
in the pattern frequency with increasing $\epsilon$.}
\end{figure}

A better measure of the transition for positive $\epsilon$
is to use the point at which the pattern becomes frequency locked to
the modulation frequency. This boundary is given by the $\times$'s
and the $+$'s in Fig. 2.
Figure 4 shows the plot of the pattern frequency as a function of 
$\epsilon$ in the absence of modulation. From this, one can see that it
is easy to distinguished the locked and unlocked patterns for
$\epsilon > 0.01$, as the frequency of the pattern differs significantly
from the Hopf frequency. With this definition of the transition,
there is no measurable hysteresis.

Figure 5 shows the effect of varying the modulation frequency for
positive $\epsilon$. As with negative $\epsilon$, the solid line
is a fit to a parabola. In this case, the parabola is centered on the
frequency of the unmodulated pattern, not the Hopf frequency. Also,
when $(f^* - f_m/2) > 0.04$, the
standing wave pattern at onset is standing squares. This is illustrated
in Fig. 6. For these images, the
modulation frequency was 0.472 Hz, and the unmodulated pattern
had a frequency of 0.305 Hz.
In this range of $(f^* - f_m/2)$,
increasing $b$ leads to a secondary transition from the standing
rectangles to standing rolls.

\begin{figure}[htb]
\epsfxsize = 3.5in
\centerline{\epsffile{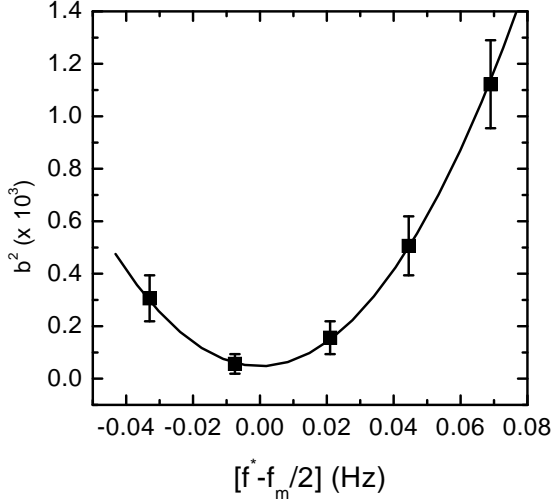}}
\caption{\label{poseps} Transition to standing waves for
$\epsilon = 0.03$. Below the curve, the system exhibits
spatiotemporal chaos. Above the curve, the system is
phase locked to the modulation frequency.
The symbols are the experimental values. The
solid line is a fit to:
$b^2 = a + b(f^*-f_m/2)^2$. In this case,
$f^* = 0.305~Hz$, which is the frequency of the pattern
at $\epsilon = 0.03$. In contrast, the Hopf frequency
is 0.34~Hz. For $|f^*-f_m/2| < 0.04$, the transition is
directly to standing rolls. For $(f^*-f_m/2) > 0.04$, the transition
is initially to standing rectangles, and there is a secondary transition
to standing rolls.}
\end{figure}

\begin{figure}[htb]
\epsfxsize = 3.5in
\centerline{\epsffile{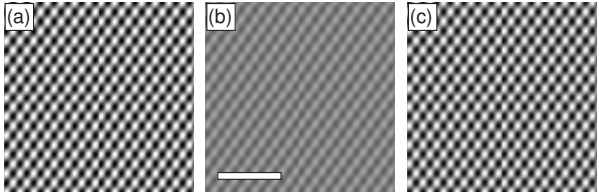}}
\caption{\label{squares} Three images from a time series
taken at $\epsilon = 0.036$, $b = 0.04$, and $f_m = 0.472$.
This corresponds to $(f^*-f_m/2) = 0.069$ in Fig.~5.
The images cover a region that is 0.6~mm x 0.6~mm and
are 0.9~s apart. The bar in (b) represents 0.2~mm.
The images have been Fourier filtered so that
only the fundamental mode is present.
These images illustrate the standing
wave nature of the pattern.}
\end{figure}

I made a qualitative survey of the behavior as a function
of the driving frequency. By increasing the driving frequency,
one decreases the angle $\theta$ between the wavevector of the pattern and
the undistorted director orientation. For our system, the same
qualitative behavior was observed for applied frequencies up
to 80~Hz. For 80~Hz, $\theta = 10^{\circ}$. For modulation at
twice the Hopf frequency, a standing roll pattern is observed. A more
detailed study of the effects of varying $\theta$ will be the
subject of future work.

The local temporal behavior of the standing roll state
is extremely regular. This is shown in
Fig. 7. Each plot is a time series of the local amplitude. The amplitude
is measured every 2 minutes, with the initial point of the
time series taken 
10 minutes after the modulation is applied. 
Figure 7a is a plot of the local amplitude as a function of 
time for $\epsilon = 0.01$ and no modulation. The amplitudes of
the right-traveling zig rolls, left-traveling zig rolls, and
left-traveling zag rolls have been shifted from their true values
by 0.015, 0.01, and -0.005, respectively. These shifts clarify the
anti-correlations present between the various modes and the irregular
variation in time. This behavior has been reported previously \cite{DAC96}.
Figure 7b is the local amplitude for
$\epsilon = 0.01$, $b = 0.02$, and $f_m = 2f_h$.
This figure shows both the regular
temporal behavior and the establishment of standing rolls (the
zag amplitude has gone to zero).

\begin{figure}[htb]
\epsfxsize = 3.5in
\centerline{\epsffile{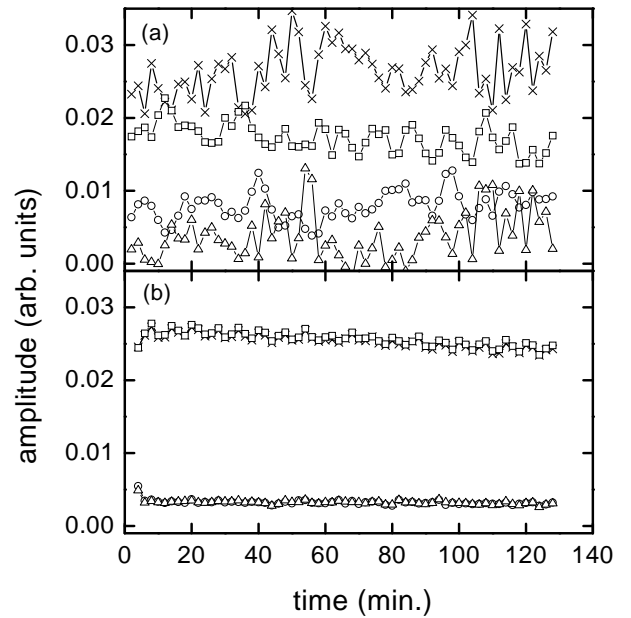}}
\caption{\label{timepow} (a) Plot of the amplitudes of
the right-traveling zig rolls ($\times$), left-traveling
zig rolls ($\Box$), right-traveling zag rolls ($\circ$),
and left-traveling zag rolls ($\triangle$). The
system was at $\epsilon = 0.01$ and $b = 0\%$. 
The amplitudes have been shifted as described in the text. (b)
Plot of the amplitudes of the right-traveling zig rolls ($\times$),
left-traveling zig rolls ($\Box$), right-traveling zag
rolls ($\circ$), and left-traveling zag rolls
($\triangle$). The system was at
$\epsilon = 0.01$ and $b = 2.0\%$.}
\label{fig:timepow}
\end{figure}

The development of the local, temporal order, generally occurred in
under a few minutes. In contrast, the spatial ordering involves
extremely long time scales. It can take up to two hours for the standing
roll domains to reach sizes on order of the system size.
However, upon removal of the modulation,
the disorder develops in a few minutes.
This difference in time scales is reflected in Fig. 2. One sees
that there is essentially no difference
between the transition to a frequency locked state and the transition to
standing waves measured by decreasing $\epsilon$. However,
standing rolls of a particular size occurred at values of $b$
slightly above the transition defined by frequency locking.
This is easily understood in terms of the
10 minute waiting time used when stepping $b$.
Clearly, the details of the spatial ordering and the multiple time scales
involved is an interesting problem. However, it is outside the scope
of this paper and will be the subject of future work.

\section{Discussion}

The transitions for negative $\epsilon$ can be directly
compared with predictions of the relevant coupled
amplitude equations \cite{RSK94}. I find excellent
agreement between the measured onset of standing
waves and the predictions of Ref.~\cite{RSK94}. The
onset to standing waves should occur when
$b = |\mu|$. Here the real part of $\mu$ is just
$\epsilon$ and the imaginary part of $\mu$ is
$2\pi\tau_d(f^* - f_m/2)$, where
$\tau_d = \gamma_1d^2/(\pi^2K_{11})$ is the director
relaxation time. Here $\gamma_1$ is a rotational
viscosity and $K_{11}$ is the splay elastic constant of
the director. For a modulation
frequency equal to twice the Hopf frequency, the
onset is given by the line $b = \epsilon$. This
is the solid line shown in Fig.~1. For the more general
case, one has $b^2 = \epsilon^2 + (2\pi\tau_d)^2(f_h - f_m/2)^2$.
The solid line in Fig. 3 is a fit to this equation. The resulting
values are: $\epsilon = -0.058 \pm 0.005$,
$\tau_d = 0.229 \pm 0.004\ {\rm s}$, and
$f_h = 0.162 \pm 0.01\ {\rm Hz}$. For comparison, the measured values of 
these parameters are: $\epsilon = -0.054 \pm 0.002$,
$\tau_d = 0.18 \pm 0.05\ {\rm s}$, and
$f_h = 0.164 \pm 0.005\ {\rm Hz}$.

The nature of the standing wave pattern depends on the nonlinear
coefficients in the amplitude equations.
The fact that I observe standing rolls at onset has important
consequences. First, the coupling coefficient between zig
and zag rolls traveling in the same direction has been
calculated \cite{TK98}. Based on this calculation, it is likely that
standing rectangles are the stable state for the parameter
range in my experiments. However, the condition for
the stability of standing rolls does involve all of the
nonlinear coefficients \cite{RSK94}, and standing rolls are not ruled
out by the calculations of Ref.~\cite{TK98}.
Therefore, these experiments highlight the need
for a determination of all of the nonlinear coefficients
before quantitative comparisons between amplitude equations
and the experiments are possible. On the other hand, in the absence
of theoretical calculations, the temporal modulation experiments
provide a means to determine the coefficients experimentally,

The results for positive $\epsilon$ are in qualitative agreement
with the predictions of Ref.~\cite{RSK94}. The critical
value of $b$ for the transition to standing waves is
linear in $\epsilon$ for fixed $f_m$.
However, I find $b = 0.577 \epsilon - 0.006$ (the solid line in Fig. 2),
and not $b = \epsilon$. This is not surprising given that 
the ground state of the experimental system is 
a state of spatiotemporal chaos. This pattern
can only be described by amplitude equations that
include spatial derivatives, and these are not included in
Ref.~\cite{RSK94}

The other qualitative agreement with Ref.~\cite{RSK94}
is the behavior as a function of modulation frequency.
The critical values of $b^2$ are
quadratic in $(f^* - f_m/2)$ for fixed $\epsilon$.
However, it is clear from Fig. 5 that $f^*$ is 
the frequency of the unmodulated pattern for the fixed
value of $\epsilon$, and not the Hopf frequency. This
is due to the shift in frequency with $\epsilon$ that is
illustrated in Fig. 4. 

An additional feature of the behavior at positive
$\epsilon$ that requires a theoretical explanation is the
regular dynamics of the standing wave state. Though an
incomplete description, the existing amplitude equation
calculations suggest that the unmodulated state is
Benjamin-Feir unstable for all wavenumbers \cite{TK98}.
This provides a possible explanation
for the spatiotemporal chaos at onset. From the
fact that the modulated state exhibits regular dynamics,
one can infer that the standing rolls are
Benjamin-Feir stable. This situation
is the opposite of that previously observed in
electroconvection in a different nematic liquid crystal \cite{JZR89}.
In that system, the
unmodulated state was stable. For high enough modulation,
the modulated state was Benjamin-Feir unstable and resulted in
irregular dynamics \cite{JZR89}. The behavior in that case agreed
well with calculations based on amplitude equations that
included the spatial derivatives\cite{F87}.
A similar
calculation is needed for the system reported on here. In particular,
it will be important to determine if temporal modulation is a
general method for eliminating spatiotemporal chaos, or if
it is specific to this system.

\acknowledgments

I thank Hermann Riecke for useful discussions. This work was
supported by NSF grant DMR-9975479.

\end{document}